\documentclass[floatfix,aps,prl,amsmath,amssymb,superscriptaddress,reprint]{revtex4-2}
\usepackage{graphicx}
\usepackage{dcolumn}
\usepackage{bm}
\usepackage{xr}
\usepackage{placeins}
\usepackage{comment}

\usepackage{ulem}%

\usepackage{float}
\newcommand{\mycomment}[1]{}
\usepackage{amsmath,graphics, setspace}
\usepackage{hyperref}
\hypersetup{
    colorlinks=true,
    linkcolor=blue,
    filecolor=magenta,      
    urlcolor=blue,
}
\usepackage[dvipsnames]{xcolor}

\newcommand{\rev}[1]{{\color{Black}#1}}
\newcommand{\revv}[1]{{\color{Black}#1}}

\begin{document}

\title{Channel flows of deformable nematics }

\author{Ioannis Hadjifrangiskou}
\affiliation{Rudolf Peierls Centre for Theoretical Physics, University of Oxford,  Oxford OX1 3PU, United Kingdom}
\author{Sumesh P. Thampi}
\affiliation{Rudolf Peierls Centre for Theoretical Physics, University of Oxford,  Oxford OX1 3PU, United Kingdom}
\affiliation{Department of Chemical Engineering, Indian Institute of Technology Madras, Chennai-36, India}
\author{Julia M. Yeomans}
\affiliation{Rudolf Peierls Centre for Theoretical Physics, University of Oxford,  Oxford OX1 3PU, United Kingdom}

\date{\today}

\begin{abstract}
We describe channel flows in a continuum model of deformable nematic particles. In a simple shear flow, deformability leads to a nonlinear coupling of strain rate and vorticity, and results in shape oscillations or flow alignment. The final steady state can depend on initial conditions, and we explain this behaviour by considering a phase space representation of the dynamics. In Poiseuille flow, particle deformability and nematic elasticity induce banding, where particles near the walls are aligned, and those near the centre of the channel oscillate in direction and shape. Our results show that particle deformability can lead to complex behaviour even in simple flows, suggesting new microfluidic experiments.

\end{abstract}

\maketitle

Deformable particles are attracting considerable interest at the intersection between soft matter and biological physics due to the emergent dynamics linked with their geometry \cite{manning2023essay}. Novel synthesis routes have made it possible to prepare and study  soft colloidal systems such as Pickering emulsions \cite{guan2024pickering}, polymer microgels \cite{scheffold2020pathways, vlassopoulos2014tunable, gnan2019microscopic} and foamy complex fluids \cite{cohen2013flow}. Softness is a characteristic feature of biological matter: MDCK cells \cite{angelini2011glass, angelini2010cell} and embryonic tissue \cite{lecuit2007cell, heisenberg2013forces, merkel2017using, blanchard2009tissue,butler2009cell} exhibit significant variations in their shape. Cell deformability is increasingly being recognized as an important aspect in cancer metastasis \cite{ahmmed2018multi, suresh2007biomechanics, wirtz2011physics}.  Recently, studies of glioblastoma in mice and humans showed that highly elongated and nematically aligned tumor cells invade the surrounding, healthy and round cells in a stream \cite{comba2022spatiotemporal} resulting in a spatial distribution of both tumour and non-tumour cells. Experiments have also demonstrated the distinct role of flow induced shear strain on the viability of cells in vasculature \cite{van2018biomimetic, boularaoui2020overview}. Further, deformability may play a role in the dynamics of self-assembled systems such as lattices of soft topological solitons observed in cholesteric liquid crystals \cite{tai2019three}. The understanding of, and ultimately, the ability to control the behaviours of such systems present novel challenges in physics and technology \cite{manning2023essay}. 

Both microscopic and macroscopic behaviours of deformable material may be traced back to their ability to change their shape, either through inter-particle interactions, or through flow induced deformations. Shape changes of individual colloids such as drops, vesicles and capsules in shear flow have been considered \cite{bagchi2009dynamics, mader2006dynamics,beaucourt2004steady}, and recent studies have focused on the development of models to describe the dynamics of motile, deformable particles \cite{tarama2014individual, poissonbracket, czajkowski2018hydrodynamics,boromand2018jamming, wang2021structural,kim2021embryonic,tarama2017dynamics, tanjeem2022shape}. However,  continuum frameworks that incorporate deformability and lead to predictions that are 
experimentally verifiable are required to further advance this field. Therefore, in this work, we use analytics and numerical simulations within a continuum model of deformable nematic particles in channel flows to reveal novel and rich dynamics. We report results for two channel flows: Couette and Poiseuille.

\textit{Model} -   Traditionally, nematic models are based on the physics of liquid crystals where the constituent particles maintain a constant, anisotropic shape \cite{onsager1949effects,maier1960einfache}. We write down coarse-grained equations of motion that describe deformable particles using the Ericksen-Leslie equations of nematic liquid crystals \cite{de1993physics}, but with a minimal extension to incorporate shape changes through the introduction of an additional variable, \rev{defined as $r = AR - 1$, where $AR$ is the aspect ratio of particles \cite{hadjifrangiskou2023active}.} In our model, the director field $\mathbf{n}$ describes the orientation of the principal axis of the deformed particles.

The governing equations that describe the evolution of the director field and the aspect ratio are 
\begin{align}
     &\left( \partial_{t} + \mathbf{u} \cdot \mathbf{\nabla} \right) n_{i} = - \lambda n_{i} + \xi\left(r \right) E_{ij}n_{j} - \Omega_{ij}n_{j} - \Gamma \dfrac{\delta \mathcal{F} }{\delta n_{i}}, \label{EL:EOM}\\
   & \left( \partial_{t} + \mathbf{u} \cdot \mathbf{\nabla} \right)r  =  2(r+1) E_{\parallel} -\Gamma_{\rev{r}}\dfrac{\delta \mathcal{F} }{\delta r} \label{omega_time_evo_main},
\end{align}
where $\mathbf{u}$ is the velocity field, $E_{ij}$ and  $\Omega_{ij}$ are the strain rate and the vorticity tensor, defined as the symmetric and anti-symmetric parts of the velocity gradient tensor $\partial_{i} u_{j}$ respectively.  $E_{\parallel} = n_{i}E_{ij}n_{j}$ is the projection of the strain rate, along the particles' orientation $\mathbf{n}$. The constant $\lambda$ is a Lagrange multiplier that guarantees that $\mathbf{n}$ is a unit vector. \revv{Eq. \eqref{omega_time_evo_main} is based on the theory of Neo-Hookean solids in a flow \cite{bilby1977finite, gao2011rheology, gao2013dynamics, hadjifrangiskou2023active}.} We take the flow alignment parameter to be a function of the \rev{shape} of the particles, $\xi(r) = \xi_{0}r$ where $\xi_{0}$ is a positive constant that sets the flow-alignment scale. The dependence of $\xi$ on $r$ is a linear approximation to a more general form derived from microscopic calculations \cite{kuzuu1983constitutive, ternet1999flow}. \revv{We neglect any coupling of $\mathbf{n}$ and $r$ back to the velocity field $\mathbf{u}$.}

$\Gamma$ and $\Gamma_{r}$ respectively set the relaxation rate of the director field and the \rev{shape variable} towards the minimum of the free energy, $\mathcal{F} = \int (f_{\rev{r}} + f_{n})\: \text{d}\mathbf{A}$, which consists of two contributions.
The first of these is associated with particle shape deformations,  $\rev{f_{\rev{r}} = \frac{A_{\rev{r}}}{2}(r - r_{0})^{2} + \frac{A_{\rev{r}}^{\star}}{4}(r - r_{0})^{4}}$, where $\rev{r_0}$ is the particle shape at equilibrium and $A_{\rev{r}}$, $A_{\rev{r}}^{\star}$ are deformation energy parameters. The second contribution, $f_n$, arising from the nematic elasticity in the one elastic constant approximation is given by $ f_{n} = \frac{K(\rev{r})}{2}\left(\nabla \mathbf{n}\right)^{2}$, where $K(\rev{r})$ is the elastic constant associated with the orientational order of the deformable particles.

\noindent\textit{Couette flow} - We first consider a continuum of deformable particles confined between two parallel plates  subjected to simple shear flow, described by the velocity profile $u_{x} = \dot{\gamma} y$. Along the channel, $\hat{x}$ direction, we assume translational invariance. The strain rate tensor for this two dimensional parallel flow, when diagonalized, reads $\frac{\dot{\gamma}}{2}\big(\begin{smallmatrix}
  1 & 0\\
  0 & -1
\end{smallmatrix}\big)$ indicating extension and compression along the principal axes. In other words an isotropic material element will deform along the principal axis of deformation, \textit{i.e.} at an angle of $\pi/4$ to the flow direction \cite{cohen2004fluid, batchelor2000introduction} in addition to the rotation induced by the vorticity tensor. 

Defining the director of the elongated particles $\mathbf{n} = \left(\cos{\theta}, \sin{\theta}\right)$, where $\theta \in (-\pi/2, \pi/2]$ is the angle to the flow direction, and non-dimensionalising time, $t' = (\Gamma_{\rev{r}}A_{\rev{r}})t$, the governing Eqs.~(\ref{EL:EOM}) and~(\ref{omega_time_evo_main}) simplify to
\begin{align}
    &\dot{r} = \alpha (r+1)\sin{2\theta} - (r-r_{0})\left(1 +  \epsilon (r-r_{0})^{2}\right), \label{r:eqn}\\
    &\dot{\theta} = \dfrac{\alpha}{2} \left(\xi_{0}r\cos{2\theta} - 1\right), \label{th:eqn}
\end{align}
where  $\alpha \equiv \dot{\gamma}/\left(\Gamma_{\rev{r}}A_{\rev{r}}\right)$ is a P\'eclet number that quantifies the relative importance of the shear to the resistance to deformation, and  $\epsilon \equiv A_{\rev{r}}^{*}/A_{\rev{r}}$ is the ratio of the elastic deformation parameters. \revv{Eqs. \eqref{r:eqn} and \eqref{th:eqn} are the long time limit of a passive nematic with a finite elastic constant subjected to a Couette flow with Neumann boundary conditions. Such a system will naturally cause all spatial gradients to vanish, hence the dynamical variables are functions of time only.} In the following analysis we use $\epsilon = 0.1$ and consider the case of inherently circular particles, \textit{i.e.,} with equilibrium shape $r_{0} = 0$.  We impose that whenever $r$ vanishes, the orientation is set to $\theta = \pi/4$.
\begin{figure}[b]
    \includegraphics[width = 8.6cm]{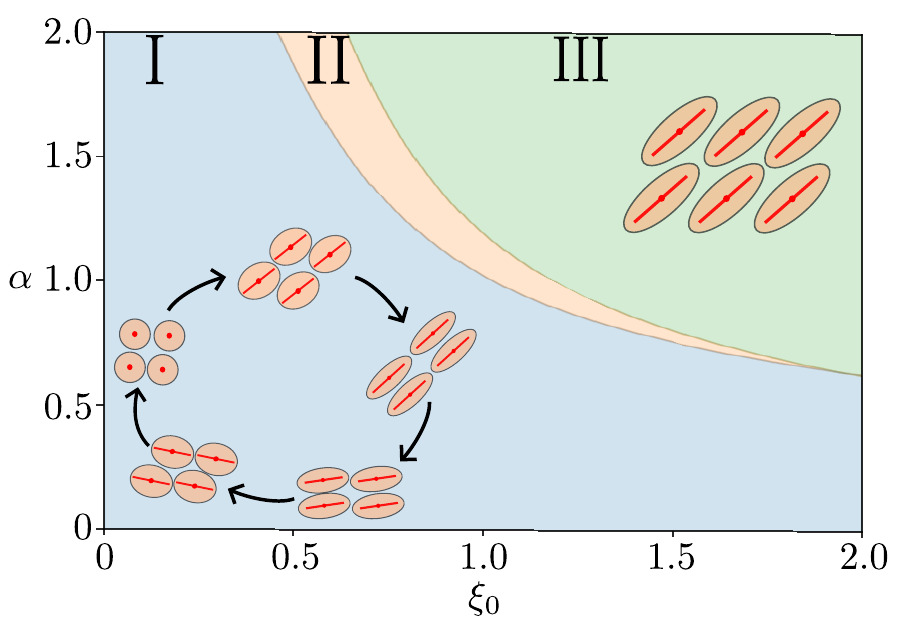}
    \caption{Dynamics of a continuum of deformable particles characterised in $\{\textnormal{P\'eclet number }(\alpha)$, $\textnormal{flow-alignment scale }(\xi_{0})\}$ parameter space. Region I: particles undergo cycles of elongation, re-orientation and contraction, III: particles align at a fixed angle to the flow, and II: both oscillatory and alignment behaviours are possible depending on initial conditions. The schematics of particles in region I and III illustrate the oscillatory and steady state behaviours.}
    \label{fig:phase_diagram}
\end{figure}

A state diagram is constructed by varying the applied shear rate, (\textit{i.e.} P\'eclet number $\alpha$) and the flow-alignment scale ($\xi_{0}$). In the case of a system of rigid particles, the strain rate and vorticity components act additively to dictate the particle kinematics,  but for the case of deformable particles, the effects due to the two components of the shear flow are coupled.
We identify three distinct regions with qualitatively different dynamics (Fig.~\ref{fig:phase_diagram}). 

\begin{figure*}
    \includegraphics[width = 17.9cm]{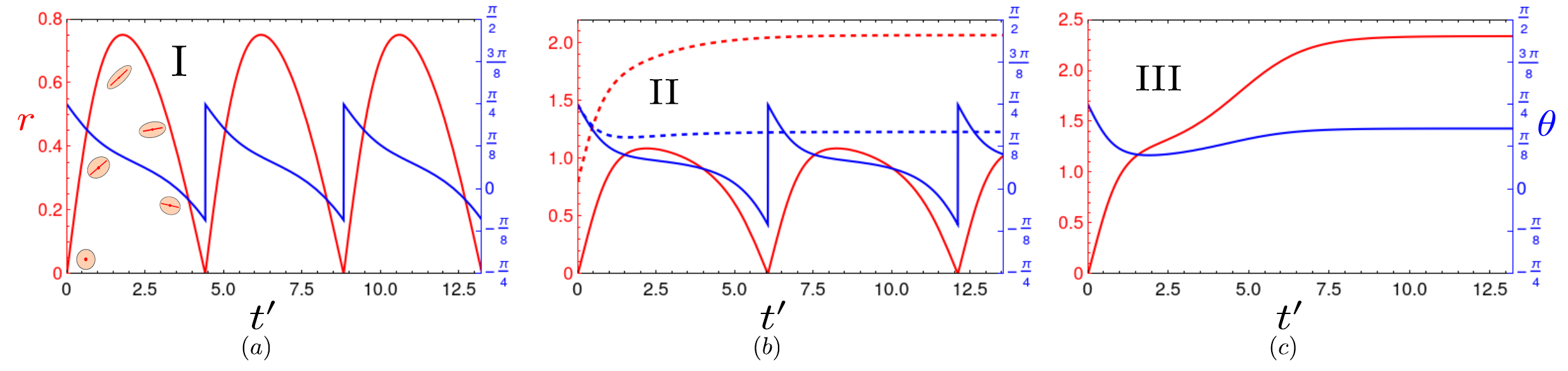}
    \caption{Time evolution of 
    $r$ (red) 
     and $\theta$ (blue) corresponding to Regions I - III. $\xi_{0} = 1$ and $\alpha = \{0.8, 1.0, 1.2\}$ in $(a)-(c)$ respectively. In $(a)$, the oscillatory evolution of the deformable particles is schematically indicated. In $(b)$ solid/dashed lines represent two different initial conditions (circular/elongated particles at $t' = 0$). 
In $(c)$, the system reaches an aligned state of deformed particles.}
    \label{fig:all_plots}
\end{figure*}

Region I corresponds to a periodic behaviour in the shape and orientation of the deformable nematic particles as shown schematically in Fig.~\ref{fig:phase_diagram} (see also Fig.~\ref{fig:all_plots}$(a)$ and movie $1$ in SM \cite{SI}). This cycle of elongation, reorientation, and contraction, emerges from the deformability of the particles. Consider circular ($r = 0$) particles at $t = 0$. 
In the absence of any inertial effects, the imposed shear flow initially elongates the particles along the principal axis of deformation $(\theta = \frac{\pi}{4})$ and $r(t)$ increases. Simultaneously the vorticity component of the shear flow rotates the particles away from the principal axis. Therefore the strain rate projected along the shape direction decreases, which results in a decrease in the elongation of the deformable particles. 
When they regain a circular shape, they are strained along the principal axis again and the cycle repeats.
The time period of these shape oscillations decreases with an increase in the shear rate, $\alpha$, and mildly increases with an increase in the flow alignment scale, $\xi_0$, (see section \rev{$3.1$} in SM \cite{SI}). We note that if $r_{0} > 0$ the particles can undergo a full tumbling motion, rotating through $2\pi$ without the shape becoming circular, (see section \rev{$4$} in SM \cite{SI}).

Region III, corresponding to flow alignment, emerges if the imposed shear rate $\alpha$ or the flow alignment scale $\xi_0$ are sufficiently large. The system reaches a steady state at an angle to the flow $\theta_s = \frac{1}{2}\arccos{\left(\frac{1}{\xi_0 r_s}\right)}$ and a particle deformation $r_s$ which increases with both $\alpha$ and $\xi_0$ (see section \rev{$3.2$} in SM \cite{SI}). Fig.~\ref{fig:all_plots}$(c)$ shows the evolution to a steady state, $r(t') = r_s, \theta(t') = \theta_s$ in region III.
The steady alignment angle $\theta_s$ reduces to the Leslie angle $\theta_L= \frac{1}{2}\arccos{\left(\frac{1}{\xi_0 r_0}\right)}$, which describes the angle that rigid nematics make with the flow axis, in the limit $\alpha \ll 1$, with $r_{0} > 0$. Deformability introduces a dependence of the alignment angle on the shear rate. Moreover, rigid nematic particles can align at any angle between $\left(0,\frac{\pi}{4}\right)$ depending upon their aspect ratio, but there exists a minimum, finite angle at which deformable nematics may align depending upon the P\'eclet number and the flow-alignment scale (see section~\rev{$3.3$} in SM \cite{SI}). 
\begin{figure}
    \centering
    \includegraphics[width = 8.6cm]{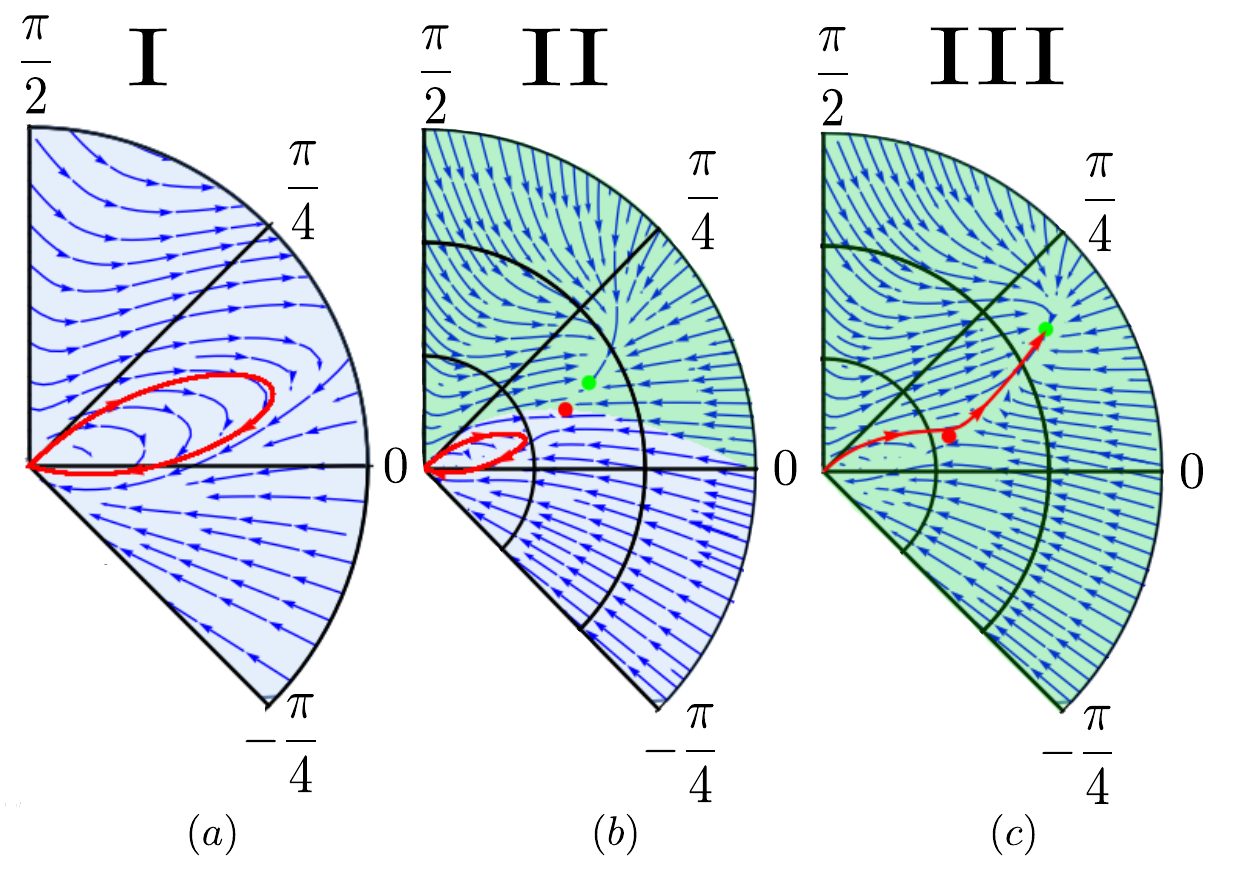}
    \caption{Phase space trajectories of the deformable nematics in $\{r,\theta\}$ space. Only a sector of the phase space is shown for brevity. Concentric circles correspond to unit intervals in $r$. Red trajectories correspond to the initial conditions $\{0, \pi/4\}$. $\xi_{0} = 1$ and $\alpha = \{0.8, 1.0, 1.2\}$ in I-III respectively. (a) Region I: all trajectories converge towards the origin, eventually joining the limit cycle indicated by the red curve. (b) Region II: steady state $\{\dot{r} = 0, \dot{\theta} = 0\}$ solutions exist as saddle (red) and stable (green) points. Green background corresponds to the attractor basin of the stable fixed point. (c) Region III: all trajectories ultimately end up at the stable fixed point. }
    \label{fig:phase_plots}
\end{figure}

\noindent The boundary separating regions I and III is not sharp. In the region labelled II in Fig.~\ref{fig:phase_diagram} the initial conditions may drive the system to a flow aligned state or to a state exhibiting shape oscillations. Fig.~\ref{fig:all_plots}$(b)$ illustrates the bistability of region II; some initial conditions lead to a steady state, whereas others result in shape oscillations.

A helpful way to clarify the distinction between the three regions is by representing the dynamics of the system in a phase space plot spanned by polar co-ordinates, $\{r, \theta\}$, as shown in Fig.~\ref{fig:phase_plots}. In the plots, the radial distance from the origin indicates the extent of deformation ($r$) while the azimuthal angle measured from the $x-$axis indicates the direction of elongation of the particles ($\theta$). The singular point at the origin corresponds to a circular particle without an attributable orientation.
The existence of the three distinct regions 
is due to the occurrence of two consecutive bifurcations that occur in the $r - \theta$ space, as the parameter $\alpha$ (or equivalently $\xi_0$) is varied. For small $\alpha$, all  phase space trajectories converge towards the origin (circular shape) and thus join the limit cycle indicated by the red curve in Fig.~\ref{fig:phase_plots}$(a)$. This limit cycle corresponds to the oscillatory shape dynamics observed in region I. As $\alpha$ is increased, a saddle-node bifurcation occurs in the phase space creating an unstable saddle point (in red) and a stable fixed point (in green) as shown in Fig.~\ref{fig:phase_plots}$(b)$, where the basin of attraction of this stable fixed point is shaded in green color. The coexistence of the limit cycle and the stable fixed point results in the bistability observed in region II. %
Finally, for sufficiently large values of $\alpha$, a homoclinic bifurcation occurs, the saddle point collides with the limit cycle and the limit cycle opens up as shown in Fig.~\ref{fig:phase_plots}$(c)$. Thus, all solutions converge towards the stable fixed point giving rise to region III. 
\revv{The equations used in the above analysis assume a constant magnitude of the nematic order parameter, which holds when flow-driven effects dominate its evolution. We discuss the relaxation of this assumption, and the effect of the nematic order parameter being time-dependent, in section 6 of the SM \cite{SI}, showing that these modifications lead only to quantitative changes in the phase diagram.}

\begin{figure*}
    \centering
    \includegraphics[width = 17.9cm]{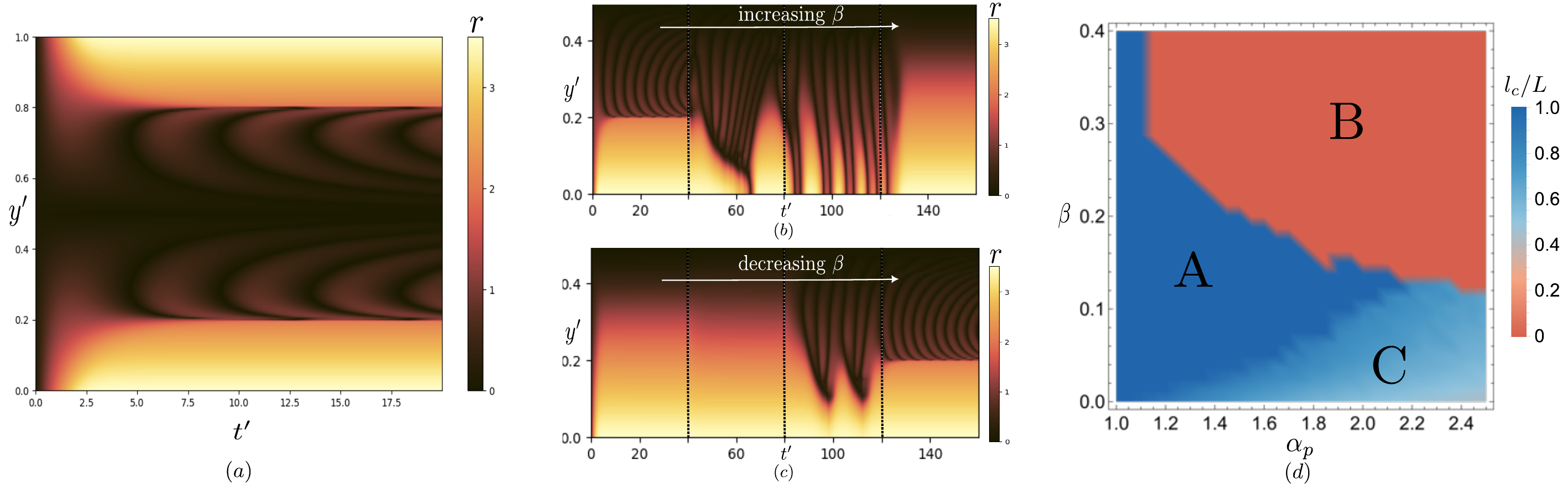}
     \caption{$(a) - (c)$ Kymographs showing spatio-temporal evolution, $r(y', t')$, of deformable nematics in a Poiseuille flow. $(a)$~$\beta = 0$. $(b)$ Increasing and $(c)$ decreasing $\beta$ between $0$ and $0.3$ in intervals of $0.1$ every 40 time steps (shown for $y' \leq 1/2$, the bottom half of the channel). In $(b)~-~(c)$ the dashed vertical lines are points in time where $\beta$ changes its value.   $(d)$ Phase diagram measuring the fraction of the central band, $l_c/L$ (defined as the portion of the system exhibiting shape oscillations) for initial conditions $r = 0.1,~\theta = \pm \pi/4$.}
    \label{fig:lplots}
\end{figure*}

\noindent\textit{Poiseuille Flow} - We now consider the continuum of deformable nematics confined between two parallel plates separated by a distance $L$, and subjected to a Poiseuille flow. 
The flow is generated by applying a pressure gradient $G$ along the length of the channel. The velocity field is given by $u_{x}(y) = \frac{G}{2\eta} y(L -y)$, \rev{where $\eta$ is the viscosity of the fluid}, with a position-dependent shear rate $\dot{\gamma}(y)$. 
The governing equations for the evolution of the elongation and the orientation of the principal axis of the deformable particles are
\begin{align}
&\dot{r} = \alpha_{p} (1 - 2y') (r + 1)\sin{2\theta} - r(1 + \epsilon r^2),\label{eq:rdotP}\\
&\dot{\theta} = \dfrac{\alpha_{p}}{2} (1 - 2y')\left(\xi_{0}r\cos{2\theta} - 1\right) + \beta^{2}\frac{r^{2}}{(r+1)^{2}}\partial_{y'}^{2}\theta, \label{eq:thetadotP}
\end{align}
where $y' = y/L \in [0,1]$ is the dimensionless spatial variable and $\alpha_{p} = {G L}/({2\eta \Gamma_{\rev{r}}A_{\rev{r}}})$ is the P\'eclet number. The last term in Eq.~(\ref{eq:thetadotP}) accounts for Frank-Oseen elasticity arising from gradients in $\theta$. The functional form of the elastic constant, $K(r) = K_{0}{r^2}/{\left(r+1\right)^2}$ is chosen as an approximation to capture the dependence on the shape of the particles \cite{tjipto1992elastic}. The dimensionless parameter $ \beta^2  = \Gamma K_{0}/(\Gamma_{\rev{r}}A_{\rev{r}}L^{2})$ controls the strength of nematic elasticity. \rev{For simplicity, other elastic terms emerging from $K(r)$ in Eqs. \eqref{eq:rdotP} and  \eqref{eq:thetadotP} are omitted. Their inclusion has no qualitative effect on the results (see section $7$ in SM  \cite{SI})}.

Simulations were run with $\alpha_{p} = 1.8, \xi_{0} = 1, \epsilon = 0.1$.  The value of $\beta$ was varied by changing $K_{0}$. Neumann boundary conditions were applied for the director field at the walls. Initial conditions were $r(y', 0) = 0.1$ and $\theta(y', 0) = \pm \pi/4$ for $y' \lessgtr 1/2$ respectively. \rev{The current analysis assumes that the imposed Poiseuille flow is unaffected by elastic stresses, and thus corresponds to the limit of large Ericksen number (see section $8$ in SM \cite{SI}).} 

In Fig.~\ref{fig:lplots}$(a)$ we plot the spatio-temporal evolution of the elongation of the deformable particles for $\beta = 0$ as a kymograph \rev{(for $\beta \neq 0$, see section 5 in SM \cite{SI})}.
The variation in the shear rate across the channel leads to particles near the walls aligning as in Region III, while particles around the center of the channel orbit as in Region I with periods that depend on the local shear rate (see movie~2 in SM \cite{SI}). Thus, (i) a wall-bound band consisting of relatively more elongated particles and (ii) a central band with relatively less elongated particles but with a spatio-temporally evolving pattern form inside the channel. 
Rigid, tumbling nematic liquid crystals can exhibit banding due to strong anchoring boundary conditions on the walls \cite{chono1998spatial,thampi2015driven},  but, here the spatial segregation into central and wall-bound bands arises purely from the deformability of the particles.

The steady state of the system is not unique for $\beta \neq 0$, but depends on the simulation protocol.
As an example, in Fig.~\ref{fig:lplots}$(b)$
we plot the spatio-temporal evolution of the  particle elongations when  $\beta$ is increased from zero to 0.3 in steps of 0.1 every 40 time steps. Results of the reverse protocol when $\beta$ is systematically decreased are shown in Fig.~\ref{fig:lplots}$(c)$. In both cases the initial conditions are ($r = 0.1,~\theta = \pm \pi/4$). The figures show qualitative differences in the dynamics for a given value of $\beta$. In particular, in the second case (Fig.~\ref{fig:lplots}$(c)$), this choice of initial conditions helps to stabilize a flow aligning state in the entire channel for large $\beta$, \rev{(see section $9$ in SM for corresponding $\theta$ kymographs \cite{SI})}.

We now quantify the state of the system as a function of the P\'eclet number, $\alpha_{p}$, and the nematic elasticity, $\beta$, for the initial conditions ($r = 0.1,~\theta = \pm \pi/4$). In the phase plot in Fig.~\ref{fig:lplots}$(d)$ the color represents $l_c/L$, the width of the central band as a fraction of the channel width:~regions A, B and C correspond to $l_c/L \approx 1, 0$ and $0\lesssim l_c/L \lesssim 1$ respectively. When $\alpha_p$ is small, even particles near the wall cannot reach a flow-aligned steady state, and hence the central band of shape oscillating particles occupies the entire channel ($l_c = L$) irrespective of the value of $\beta$  (Region A). With an increase in $\alpha_p$, and at $\beta \lesssim 0.2$ there is flow-alignment near the walls,  decreasing the width of the central band  (Region C). For larger $\beta$ (see also section \rev{$10$} in SM \cite{SI}) the system is not able to overcome the nematic initial conditions and all nematogens remain flow-aligned (Region B).  However, Region B disappears for random initial conditions.

To summarize, our study of Couette flow establishes a framework to understand the coupled dynamics of shape and orientation in a continuum of deformable particles. Results for Poiseuille flow show how more complex solutions emerge due to shape deformations and nematic elasticity. In both cases, our model gives results that contrast with nematic models of rigid particles, and also lend themselves to experimental verification.  \rev{In the SM \cite{SI} we consider the presence of additional elastic terms and removing the assumptions of infinite Ericksen number and constant magnitude of the nematic order. No qualitative differences are seen. Thus, the simple model proposed here is a useful way of analyzing the dynamics and rheology of deformable nematics.} 
Further, many eukaryotic cells are deformable and extension of the current analysis for cellular layers and tissues would be interesting.
\\
\begin{acknowledgments}
IH would like to acknowledge funding from the Gould \& Watson Scholarship. SPT thanks the Royal Society and the Wolfson Foundation for the Royal Society Wolfson Fellowship award and acknowledges the support of the Department of Science and Technology, India via the research grant CRG/2023/000169. JMY acknowledges support from the ERC Advanced GrantActBio (funded as UKRI Frontier Research GrantEP/Y033981/1). The authors would like to thank R. Valani, J. Rozman and S. Bhattacharyya for helpful discussions.
\end{acknowledgments}

\bibliography{references} 

-----------------------------------------

\end{document}